\documentclass[conference]{IEEEtran}
\IEEEoverridecommandlockouts
\usepackage{cite}
\usepackage{amsmath,amssymb,amsfonts}
\usepackage{algorithmic}
\usepackage{graphicx}
\usepackage{textcomp}
\usepackage{xcolor}

\usepackage{amsmath, amssymb, amsfonts, amsthm, booktabs, cancel, comment, enumitem, eurosym, float, graphicx, mathtools, multirow, nicefrac, placeins, soul, steinmetz, stfloats, siunitx, svg, url, xcolor, tabularx}%
\usepackage{commath}
\usepackage{cite}
\usepackage{subfigure}
\usepackage{hyperref}
\usepackage{nomencl}
\usepackage{cleveref}
\makenomenclature

\floatstyle{ruled}
\newfloat{model}{thp}{lop}
\floatname{model}{\textsc{Model}}

\makeatletter
\renewcommand*{\fnum@model}{\fname@model}
\makeatother

\def\BibTeX{{\rm B\kern-.05em{\sc i\kern-.025em b}\kern-.08em
    T\kern-.1667em\lower.7ex\hbox{E}\kern-.125emX}}
\begin{document}

\bstctlcite{IEEEexample:BSTcontrol}

\title{Assessing Distribution Network Flexibility via Reliability-based P-Q Area Segmentation

\thanks{This work was carried out as a part of the ATTEST project (the Horizon 2020 research and innovation programme, grant agreement No 864298).}
}

\author{
\IEEEauthorblockN{Andrey Churkin\textsuperscript{1}, Wangwei Kong\textsuperscript{1}, Jose N. Melchor Gutierrez\textsuperscript{1},\\ Pierluigi Mancarella\textsuperscript{1,2}, Eduardo A. Martínez Ceseña\textsuperscript{1,3}}
\vspace{1\jot}
\IEEEauthorblockA{\textsuperscript{1}Department of Electrical and Electronic Engineering, The University of Manchester, Manchester, UK\\
\textsuperscript{2}School of Electrical and Electronic Engineering, The University of Melbourne, Melbourne, Australia\\
\textsuperscript{3}Tyndall Centre for Climate Change Research, UK\\
\{andrey.churkin, p.mancarella, alex.martinezcesena\}@manchester.ac.uk}
}


\maketitle

\begin{abstract}

This paper proposes a framework to assess the flexibility of active distribution networks (ADNs) via  P-Q area segmentation, considering the reliability of flexible units (FUs). A mixed-integer quadratically constrained programming (MIQCP) model is formulated to analyse flexible active and reactive power support at the interface with transmission networks, explicitly capturing the contributions and reliability of FUs that provide flexibility services within an ADN. The numerical simulations performed for a real 124-bus UK distribution network demonstrate the optimal flexibility provision by different FUs, as well as the corresponding reliability and the impact of network reconfiguration. Distribution system operators (DSOs) can use the proposed framework to identify critical units, select an adequate combination of flexibility volumes, and manage its reliability.
\end{abstract}

\begin{IEEEkeywords}
Active distribution network, combinatorial optimisation, flexibility services, reliability, TSO-DSO coordination.\end{IEEEkeywords}

\section{Introduction}
The emergence of distributed energy resources that can act as flexible units (FUs), coupled with information and communications technologies and intelligent controls, brings new opportunities to enhance the efficiency and flexibility of distribution networks operation \cite{Eid2016}. In this context, the increased controllability of the now active distribution networks (ADNs) allows managing power injections at specific points (e.g., at a substation or a feeder), making ADNs natural providers of flexibility services such as active and reactive power support and voltage control \cite{Chowdhury2009}. This has motivated research on the flexibility potential across entire distribution networks, which would allow distribution system operators (DSOs) to trade active and reactive power with transmission system operators (TSOs). Such arrangements enable DSOs to use part of their flexibility to support the distribution network, whereas trading the rest at the TSO/DSO interface to support the transmission network (e.g., in congestion management) and the overall power system operation (via ancillary services). 

Multiple coordination schemes have recently been proposed to enable the provision of flexibility services between DSOs and TSOs \cite{Vicente-Pastor2019,Edmunds2020}. Flexibility services provided by ADNs can be characterised by an aggregated P-Q flexibility area, defined as feasible combinations of active and reactive power exchanges at the TSO/DSO interface \cite{Riaz2021}. In studies such as \cite{Gonzalez2018,Riaz2019}, Monte Carlo simulations were used to map the availability and cost of flexible power at the TSO/DSO interface. This approach requires a large number of randomly generated operating points, some of which can be infeasible and, therefore, get discarded. Despite their simplicity, such methods suffer from significant computational burden and approximation errors. To overcome these limitations, more advanced flexibility area assessment techniques were developed, such as \cite{Silva2018,Contreras2018,Capitanescu2018}. These techniques aim to approximate the boundary of the feasibility area by considering FUs limits and network constraints. Thus, the ADN flexibility can be estimated through a lower number of targeted simulations. Additionally, these algorithms allow explicit control over the accuracy of the approximation, which can be set as the desired level of the area’s granularity.

In existing approaches for P-Q flexibility areas estimation, it is generally assumed that FUs are fully available for flexibility services provision. 
This might not be reasonable in practical applications, as the reliability of each FU to deliver the full requested P-Q output is uncertain. For example, some units may not be available for flexibility services provision or may fail to provide estimated flexible power support. However, few attempts have been made to characterise the structure of ADN flexibility and incorporate the contributions and uncertainties associated with FUs activation. To capture the uncertainties of ADN flexibility, study \cite{Stankovic2020} proposed probabilistic reactive power capability charts that represent reactive power support limits of distribution systems as families of random variables with their associated probabilistic density functions. This approach identifies different levels of flexibility with associated uncertainties, but the structure of ADN flexibility and the contributions of each FU are concealed. ADN flexibility can also be modelled by solving a set of scenario-based robust optimisation problems \cite{Kalantar-Neyestanaki2020,Tan2020}. However, such approaches require large data sets of scenarios to incorporate uncertainties.

The lack of the explicit components of ADN flexibility areas limits the capability to request services with desired levels of reliability (e.g., TSO might request the maximum flexible power output that can be provided with a 95\% confidence level). Understanding the contribution of each FU to the flexibility areas is also relevant when delivering services that require specific metrics, such as dynamic (ramp) flexibility, duration, cost, etc. \cite{Riaz2021}. Furthermore, it is crucial to perform disaggregation of ADN flexibility to obtain dispatching commands for FUs during real-time operation \cite{Yi2021}.

To address the mentioned research gaps, this paper proposes a new approach to characterise the flexibility services provided by ADNs based on P-Q flexibility area segmentation. For this purpose, a combinatorial optimisation model is formulated to classify the segments of the flexibility area by the number of FUs activated and the reliability associated with units activation. This approach explicitly models the contributions of each FU and requires little information to build portfolios of ADN flexibility, which makes it suitable for practical applications. The resulting classification provides detailed information for DSOs to manage uncertainties associated with the use of flexibility and trade ancillary services with the desired level of reliability. Moreover, flexibility P-Q area segmentation captures the effects of network reconfiguration and changes in the reliability estimation of FUs activation.

The proposed framework is tested on a 124-bus radial distribution network from the UK, which can be reconfigured to meet reliability standards. The results demonstrate that the contributions and reliabilities of FUs across segments of the ADN flexibility area vary drastically, especially subject to their locations and electrical distance to the TSO/DSO interface.

\section{Methodology}
In this section, a mixed-integer quadratically constrained programming (MIQCP) formulation is proposed to explicitly map the operation of every FU deployed to provide P-Q support. The methodology includes an optimal power flow (OPF) model and dedicated algorithms for aggregating flexibility, estimating its limits, and segmenting the P-Q flexibility areas based on the number of FUs activated and their reliability.

\subsection{Combinatorial Optimal Power Flow}
A variety of OPF models can be used to identify feasible operating points of FUs within an ADN and produce P-Q flexibility areas. 
In this work, the DistFlow OPF model \eqref{Model1: objective}-\eqref{Model1: voltage limits} is selected, which enables formulating network constraints as quadratic equations \cite{Baran1989-1,Baran1989-2}.\footnote{Note that the DistFlow model is equivalent to the exact AC power flow equations, but is only valid for radial networks, such is the case study analysed in Section~\ref{Section: case study}. Other models can be selected to handle more complex network configurations \cite{Silva2018,Capitanescu2018}.}
The model defines active and reactive power flows through each branch, $p_{ij}$, $q_{ij}$, as a function of generator outputs, $p_{i,g}$, $q_{i,g}$, nodal voltages, $v_i$, and branch currents, $i_{ij}$. The formulation is simplified by substituting the products of the nodal voltage variables by $w_i$ and the squared branch currents by $l_{ij}$, i.e., $v_i^2=w_i$, $i_{ij}^2=l_{ij}$. The DistFlow model is modified by including power outputs of FUs, $p_{i,f}$, $q_{i,f}$, and corresponding binary variables to map the activation of each FU, $x_{i,f}$.

\begin{model}[t]
\caption{Modified DistFlow OPF \hfill [MIQCP]}
\label{TEP1}
\begin{subequations} 
\label{Mod: TEP1}
\vspace{-3\jot}
\begin{IEEEeqnarray}{lll}
    {\textbf{Variables:}} & \IEEEnonumber\\
    p_{i,g}, q_{i,g}  &\forall i \in \mathcal{N}, g \in \mathcal{G} & \IEEEnonumber \\
    p_{ij}, q_{ij}  &\forall (i,j) \in \mathcal{L} & \IEEEnonumber \\
    v_i \quad\, ({v_i}^2=w_i) &\forall i \in \mathcal{N} & \IEEEnonumber \\
    i_{ij} \quad ({i_{ij}}^2=l_{ij})  &\forall (i,j) \in \mathcal{L} & \IEEEnonumber \\
    p_{i,f}, q_{i,f}  &\forall i \in \mathcal{N}, f \in \mathcal{F} & \IEEEnonumber \\
    x_{i,f} \in \{0,1\}  &\forall i \in \mathcal{N}, f \in \mathcal{F} & \IEEEnonumber \\
    {\textbf{Objective:}} & \IEEEnonumber\\
     \min \enskip \pi_p \smashoperator{\sum_{(i,j) \in \mathcal{L}}} p_{ij} + \pi_q \smashoperator{\sum_{(i,j) \in \mathcal{L}}} q_{ij} &
         \begin{aligned}
        &i=i^\text{ref}\\
        &\pi_p, \pi_q \in\\
        & \{-1,0,1\}
        \end{aligned}
    \quad
    \label{Model1: objective} \IEEEyesnumber\\
    {\textbf{Constraints:}} & \IEEEnonumber\\
        p^{\text{min}}_{i,g} \leq p_{i,g} \leq p^{\text{max}}_{i,g} &\forall i \in \mathcal{N}, g \in \mathcal{G} \qquad\quad \label{Model1: Pgen. output} \\
        q^{\text{min}}_{i,g} \leq q_{i,g} \leq q^{\text{max}}_{i,g} &\forall i \in \mathcal{N}, g \in \mathcal{G} \label{Model1: Qgen. output} \\
        p_{i,f},q_{i,f} \in \mathcal{S}_{i,f} &\forall i \in \mathcal{N}, f \in \mathcal{F} \label{Model1: flex. output} \\
        p_{ij} = p_{j,d} - p_{j,g} - x_{j,f} p_{j,f} \label{Model1: Pflow}\\
        \qquad + \Re(Z_{ij})l_{ij} + \smashoperator{\sum_{(j,k) \in \mathcal{L}}}p_{jk} \quad &\forall (i,j) \in \mathcal{L}   \IEEEnonumber\\
        q_{ij} = q_{j,d} - q_{j,g} - x_{j,f} q_{j,f} \label{Model1: Qflow}\\
        \qquad + \Im(Z_{ij})l_{ij} + \smashoperator{\sum_{(j,k) \in \mathcal{L}}}q_{jk} \quad &\forall (i,j) \in \mathcal{L} \IEEEnonumber\\
        w_j = w_i + \abs{Z_{ij}}^2 l_{ij} \label{Model1: voltage relation}\\
        \qquad - 2\big{(}\Re(Z_{ij})p_{ij} + \Im(Z_{ij})q_{ij}\big{)} \qquad &\forall (i,j) \in \mathcal{L} \IEEEnonumber  \\
        p_{ij}^2 + q_{ij}^2 = l_{ij}w_i  &\forall (i,j) \in \mathcal{L} \quad \label{Model1: apparent power} \\
        p_{ij}^2 + q_{ij}^2 \leq (S_{ij}^\text{max})^2  &\forall (i,j) \in \mathcal{L} \quad \label{Model1: line limits} \\
        (v_i^\text{min})^2 \leq w_i \leq (v_i^\text{max})^2  &\forall i \in \mathcal{N} \quad \label{Model1: voltage limits}
    \vspace{-1\jot}
\end{IEEEeqnarray}
\end{subequations}
\end{model}

To produce P-Q flexibility areas for a selected \textit{reference} location (e.g., TSO/DSO interface), the objective function \eqref{Model1: objective} is iteratively updated to minimize or maximize the distribution network's power consumption at the reference node, $i^\text{ref}$. For this purpose, the coefficients $\pi_p$ and $\pi_q$ are introduced as a means to control the optimisation direction. Constraints \eqref{Model1: Pgen. output} and \eqref{Model1: Qgen. output} limit active and reactive power outputs of generators. Constraint \eqref{Model1: flex. output} defines the flexible capacity of each FU as a P-Q capability set $\mathcal{S}_{i,f}$, such that $p_{i,f}$ and $q_{i,f}$ are feasible flexible power outputs available at node $i$. Constraints \eqref{Model1: Pflow} and \eqref{Model1: Qflow} define active and reactive power flows between nodes $i$ and $j$. For each node $j$, net nodal active and reactive power withdrawals are given by the differences between power demands, $p_{j,d}$, $q_{j,d}$, generator outputs, $p_{j,g}$, $q_{j,g}$, FU outputs, $p_{j,f}$, $q_{j,f}$ (each coupled with a binary decision variable $x_{j,f}$), and power flows via connected lines, $p_{jk}$ and $q_{jk}$. Active and reactive power losses in each line are represented by $\Re(Z_{ij})l_{ij}$ and $\Im(Z_{ij})l_{ij}$, where $Z_{ij}$ is branch impedance.
The voltage relation between nodes $i$ and $j$ is given by \eqref{Model1: voltage relation}. Constraint \eqref{Model1: apparent power} defines the relation between power flows and currents. The apparent power of each line and the voltages are limited with \eqref{Model1: line limits} and \eqref{Model1: voltage limits}, respectively.

\subsection{Flexibility Aggregation}
The aggregated ADN flexibility area describing all feasible TSO-DSO active and reactive power exchanges can be found by solving a sequence of OPF problems \eqref{Model1: objective}-\eqref{Model1: voltage limits}. In this work, the $\varepsilon$-constraint method is selected to approximate P-Q flexibility areas at a selected reference bus, i.e., TSO/DSO interface \cite{Capitanescu2018}. This approach offers greater control over the accuracy of the flexibility area than alternative approaches such as random sampling \cite{Riaz2019} and radial reconstruction methods \cite{Contreras2018}. Moreover, the $\varepsilon$-constraint method guarantees that the extreme points of the flexibility area are identified. The first step of the method involves estimating the extreme points of the flexible area, e.g., maximum and minimum active and reactive power exchanges at the TSO/DSO interface. Then, the identified ranges of flexible active and reactive power are divided into a selected number of $k$ intervals, which dictates the accuracy of the model. For each interval, components of aggregated flexibility can be limited by using inequality constraints \eqref{Model boundary: epsilon p} or \eqref{Model boundary: epsilon q}, where $\varepsilon$ is a small positive constant.\footnote{It is important to choose an appropriate value for $\varepsilon$ that is small enough to accurately limit the P and Q components of aggregated flexibility, but large enough to ensure numerical stability, feasibility, and solver convergence.}
\begin{subequations} 
\begin{IEEEeqnarray}{lll}
    P_{i}^k - \varepsilon^p \leq {\sum_{(i,j) \in \mathcal{L}}} p_{ij} \leq  P_{i}^k + \varepsilon^p \qquad & i = i^{\text{ref}} \label{Model boundary: epsilon p} \label{Interval P}\\
    Q_{i}^k - \varepsilon^q \leq {\sum_{(i,j) \in \mathcal{L}}} q_{ij} \leq  Q_{i}^k + \varepsilon^q \qquad & i = i^{\text{ref}}\label{Model boundary: epsilon q} \label{Interval Q}
\end{IEEEeqnarray}
\end{subequations} 
For example, if considering $k$ intervals between maximum and minimum aggregated reactive power, constraint \eqref{Model boundary: epsilon q} is added to limit the reactive power at interval values $Q_{i}^k$. Then, for each interval, model \eqref{Model1: objective}-\eqref{Model1: voltage limits} is solved two times to find the minimum and maximum active power exchanges. Thus, by dividing the flexibility area into $k$ intervals, the model approximates it with $2k$ boundary points.
 
\subsection{P-Q Flexibility Area Segmentation by the Number of FUs}
The binary variables $x_{i,f}$ included in model \eqref{Model1: objective}-\eqref{Model1: voltage limits} are used to explicitly capture the contribution of each FU to the aggregated ADN flexibility at the reference node (TSO/DSO interface). In this work, the full available capacity of each FU is considered for the provision of flexibility services. However, this assumption is optimistic because, in practice, the volume of power that can be delivered at the reference node will likely be lower as some FUs may be unavailable, may have committed part of their capacity to other services, and so on. Accordingly, the proposed model can be improved with robust/chance constraints and other less optimistic approaches \cite{Stankovic2020,Kalantar-Neyestanaki2020,Tan2020}, which is the subject of future research.

The structure of the P-Q flexibility area and individual contributions by different FUs can be analysed by dividing the area into segments. In each segment, the number of activated FUs is limited using the following constraint:
\begin{IEEEeqnarray}{lll}
    \label{Flex. number limit}
    \smashoperator{\sum_{\forall i \in \mathcal{N}, f \in \mathcal{F}}} x_{i,f} \leq {X}_{\text{seg}} \qquad \forall \text{seg} \in S
\end{IEEEeqnarray}


In this manner, it is possible to map the aggregated P-Q flexibility area with $|S|$ segments based on the number of activated FUs. The key advantage of this mapping approach is that it facilitates identifying the FUs that are required for each P-Q flexibility service. This information can be used to select the lowest number of FUs that offer a certain service request (e.g., from the TSO). As will be further discussed below, this segmentation approach also facilitates customising the reliability levels of the flexibility services. 

\subsection{Reliability-based Segmentation of P-Q Flexibility Area}
To demonstrate the reliability-based segmentation of P-Q flexibility areas, in this work, the reliability of each FU is described by a coefficient $R_{i,f}$. Under this assumption, the reliability of a flexibility segment can be calculated as the product of the coefficients for activated FUs:
\begin{IEEEeqnarray}{lll}
    \label{Flex. reliability}
    \mathcal{R}_{\text{seg}} = \prod_{\forall i \in \mathcal{N}, f \in \mathcal{F} | x_{i,f} = 1} R_{i,f} \qquad\quad \forall \text{seg} \in S
\end{IEEEeqnarray}

This approach captures that, as expected, reliability decreases with the number of FUs required for the flexibility services. The services at the boundary of the aggregated P-Q flexibility area are the least reliable as all FUs would have to be activated. Accordingly, the model can rank segments of the P-Q flexibility areas based on their reliabilities, which can be used by DSO to manage the volumes and reliability of the flexibility services traded at the TSO/DSO interface.

\section{Case Study}\label{Section: case study}
\begin{figure*}
    \centering
    \includegraphics[width=0.9\textwidth]{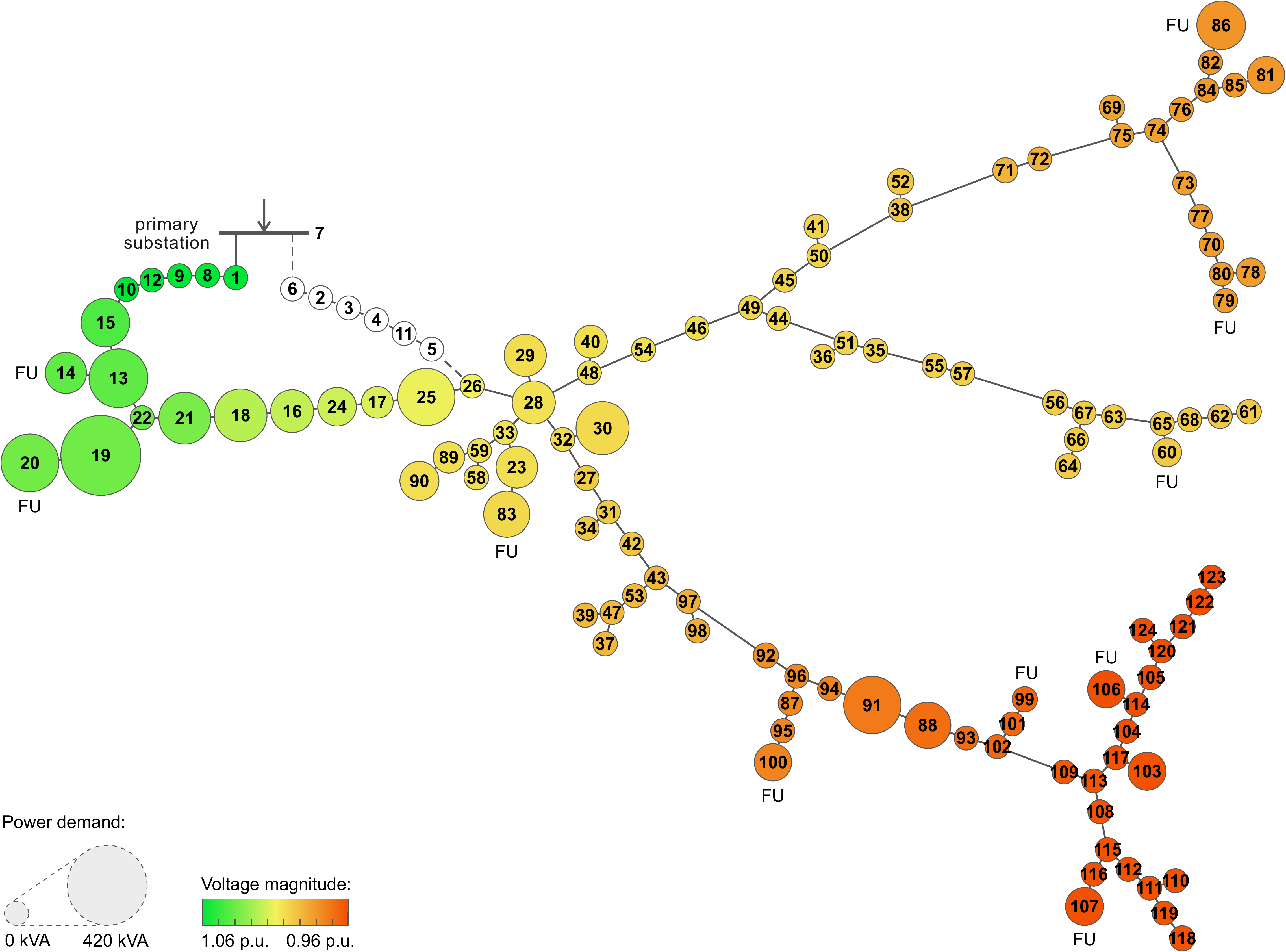}
    \caption{Case study: 124-bus radial distribution network with 10 FUs.}
    \label{Fig: MLData_Griffin case - scheme}
\end{figure*}

The proposed methodology is demonstrated with a real 124-bus 6.6 kV distribution network in the UK \cite{network_UK}. The network is visualised as a graph in Fig.~\ref{Fig: MLData_Griffin case - scheme} using the \textit{ForceAtlas2} algorithm \cite{Jacomy2014}. The sizes of the nodes represent the power demand at each bus. The lengths of the edges are proportional to the impedance of the lines. Nodes are colored according to their voltage levels during reference conditions, i.e., assuming that all FUs are switched off. This highlights areas with high or low voltages which may limit the operation of FUs. The network can be reconfigured to supply the demand through either branches 7-1 or 7-6. In the figure, the network is supplied via feeder 7-1 (path 7-26 is displayed with dashed lines).\footnote{Note that nodes 2-6 and 11 are switching stations with no loads. Therefore, they are displayed disconnected in Fig.~\ref{Fig: MLData_Griffin case - scheme}, as this simulation corresponds to supply via feeder 7-1.} The system is typically configured as radial feeders, and reconfiguration is only used to reconnect customers and meet reliability standards.

\begin{figure*}
    \centering
    \subfigure(a){\includegraphics[width=0.47\linewidth]{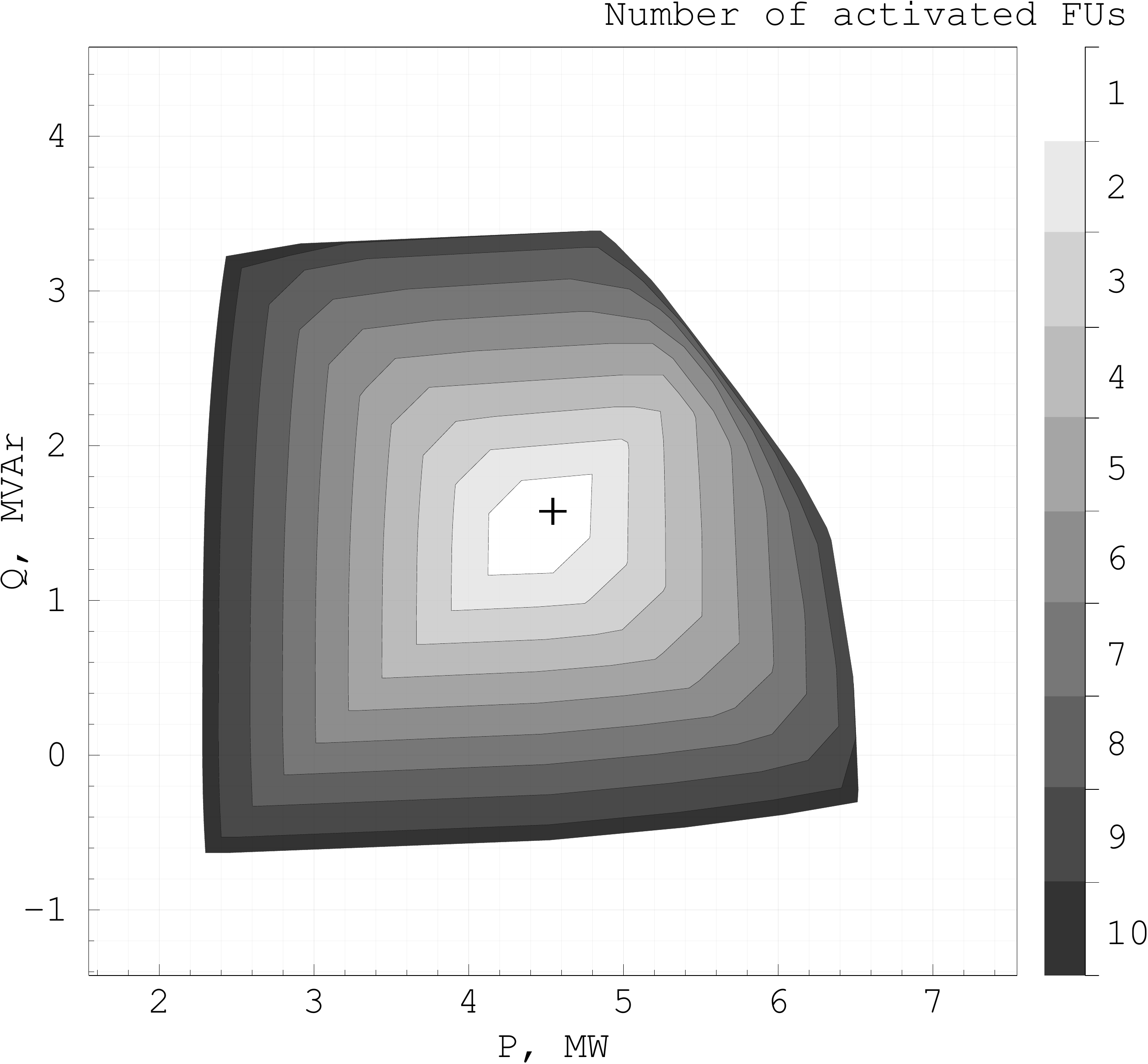}}
    \subfigure(b){\includegraphics[width=0.47\linewidth]{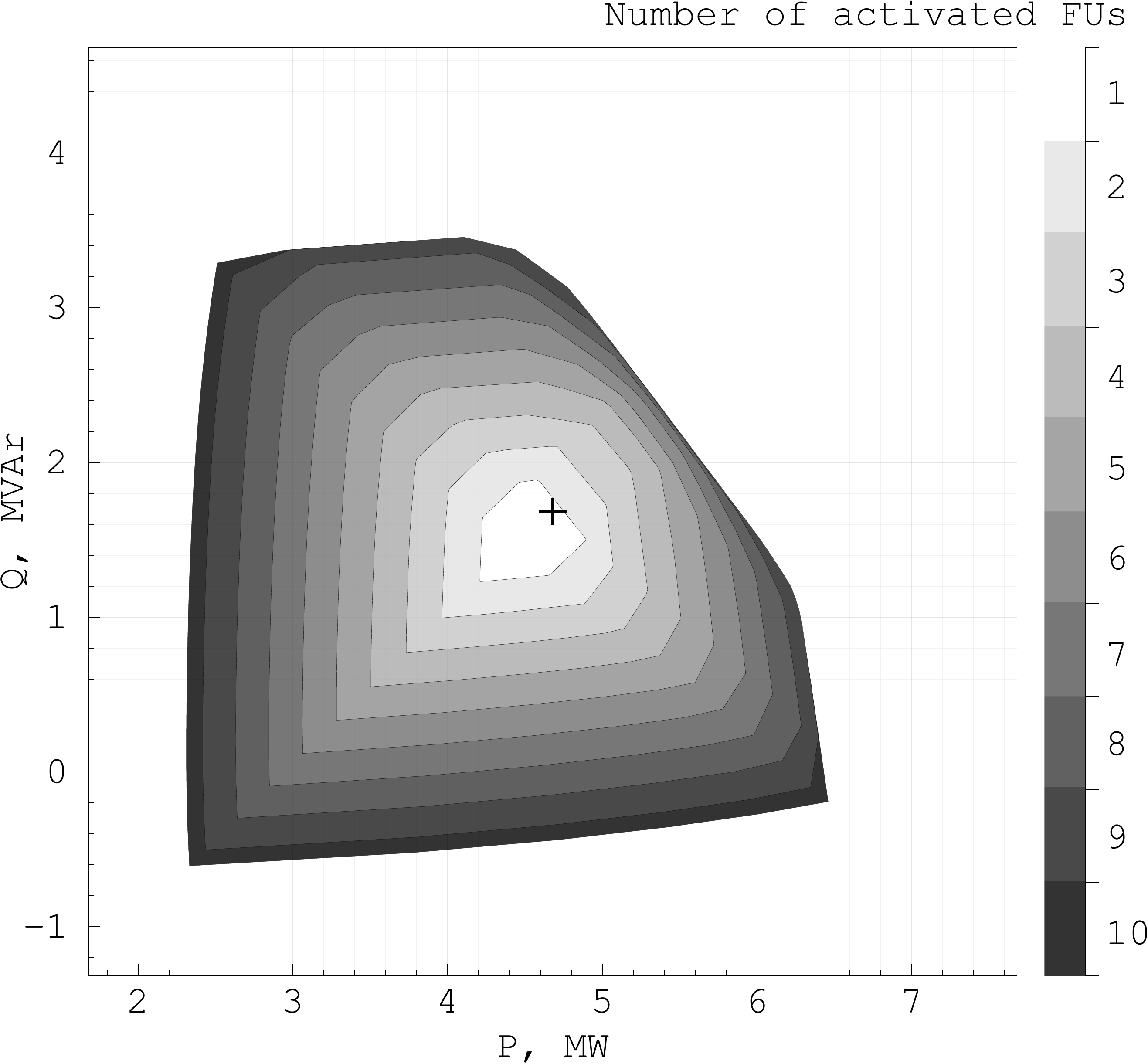}}
    \caption{Flexibility P-Q area segmentation for the 124-bus network based on the number of FUs activated: (a) network configuration with the power supply via feeder 7-1; (b) supply via feeder 7-6. The grey color scheme indicates the number of FUs activated in each of the segments. 
    The markers correspond to the initial operating point of the network (all FUs are switched off), while the coordinates represent the network’s power consumption.
    }
    \label{Fig: Griffin - segments by units}
\end{figure*}

As shown in Fig.~\ref{Fig: MLData_Griffin case - scheme}, ten FUs are located in the network. The P-Q capabilities and reliability of each unit are given in Table~\ref{table: data}, where the negative and positive regulation limits correspond to flexible power consumption and generation. The proposed MIQCP model \eqref{Model1: objective}-\eqref{Model1: voltage limits} was used to characterise the contributions of the FUs and build flexibility area segmentation by the number of units activated. The resulting P-Q area segmentation is presented in Fig.~\ref{Fig: Griffin - segments by units}. The MIQCP model was implemented in JuMP 0.21.8 for Julia 1.6.1 programming language and solved with the Gurobi 9.1.2 solver. A laptop with an Intel Core i7-10510U CPU 1.80GHz and 16 GB of RAM was used for the simulations. To estimate the aggregated flexibility areas and perform their segmentation, the model was solved iteratively for 50 $\varepsilon$-constrained intervals (i.e., the boundary of each segment was approximated by 100 points), and the average computational time to produce the P-Q area segmentation for this case study was 970 seconds.

\begin{table}
\renewcommand{\arraystretch}{1.1}
\caption{Parameters of Flexible Units Placed in the Network}
\centering
\begin{tabular}{llll}
\toprule 
FU location & P regulation limits  & Q regulation limits  & Reliability\\
(bus \#) & (MW) & (MVAr) & $R_{i,f}$ \\
\midrule 
14 & [-0.2,0.2] & [-0.2,0.2] & 0.970\\ 
20 & [0.0,0.4] & [0.0,0.4] & 0.990\\
83 & [-0.2,0.2] & [-0.2,0.2] & 0.975\\
60 & [-0.1,0.1] & [-0.1,0.1] & 0.965\\ 
79 & [-0.2,0.2] & [-0.2,0.2] & 0.960\\
86 & [-0.2,0.2] & [-0.2,0.2] & 0.980\\
99 & [-0.2,0.2] & [-0.2,0.2] & 0.955\\ 
100 & [-0.2,0.2] & [-0.2,0.2] & 0.950\\ 
106 & [-0.2,0.2] & [-0.2,0.2] & 0.945\\
107 & [-0.2,0.2] & [-0.2,0.2] & 0.985\\
\bottomrule 
\end{tabular}
\label{table: data}
\end{table}

The segmentation presented in Fig.~\ref{Fig: Griffin - segments by units} identifies the P-Q limits of the aggregated network flexibility that can be reached by activating different numbers of units.
Since the FUs under consideration have different P-Q capabilities, the shapes of the segments are determined by the combinations of their capability sets, subject to network constraints. The segments do not expand equally in all directions. For example, the central segment is formed by the contributions of unit 20 (FU with the highest capacity) and units 106 and 107 (the most distant units that can increase the network's consumption and power losses). As more units get activated, the voltage regulation limits set by $v_i^\text{min}=0.94$ p.u. and $v_i^\text{max}=1.06$ p.u. restrict the outputs of some FUs. This effect can be observed in the right upper side of the plot, where diminishing contributions of additionally activated units occur. Fig.~\ref{Fig: Griffin - segments by units} also highlights that the flexibility areas and their reliability can change significantly under different network configurations.

In the performed segmentation based on the number of units activated, all FUs were considered equally reliable and available for flexibility services provision. However, in practice, the reliability of each FU may vary. This happens due to the uncertainties associated with the availability of different technologies and forecasts for FUs generation. There can be a lack of data from electric vehicle aggregators, the availability of consumers participating in demand response programs can be uncertain, etc. Considering such factors, DSOs can measure the level of reliability of each FU and manage the network's flexibility by activating units with high reliability of providing flexible power. To illustrate this, different reliabilities $R_{i,f}$ were assigned to the FUs located in the network, as listed in Table~\ref{table: data}. The unit at node 20 is considered the most reliable one with the coefficient $R_{20}=0.99$. Other units have reliability between $0.945$ and $0.985$. Under these assumptions, each unit is reliable enough to provide flexible power. However, the aggregated flexibility of the network is formed by different combinations of FUs. The resulting reliability of a segment can be significantly lower than the reliability of individual units. For example, the reliability of simultaneous activation of all ten FUs is $0.718$. 

\begin{figure*}
    \centering
    \subfigure(a){\includegraphics[width=0.47\linewidth]{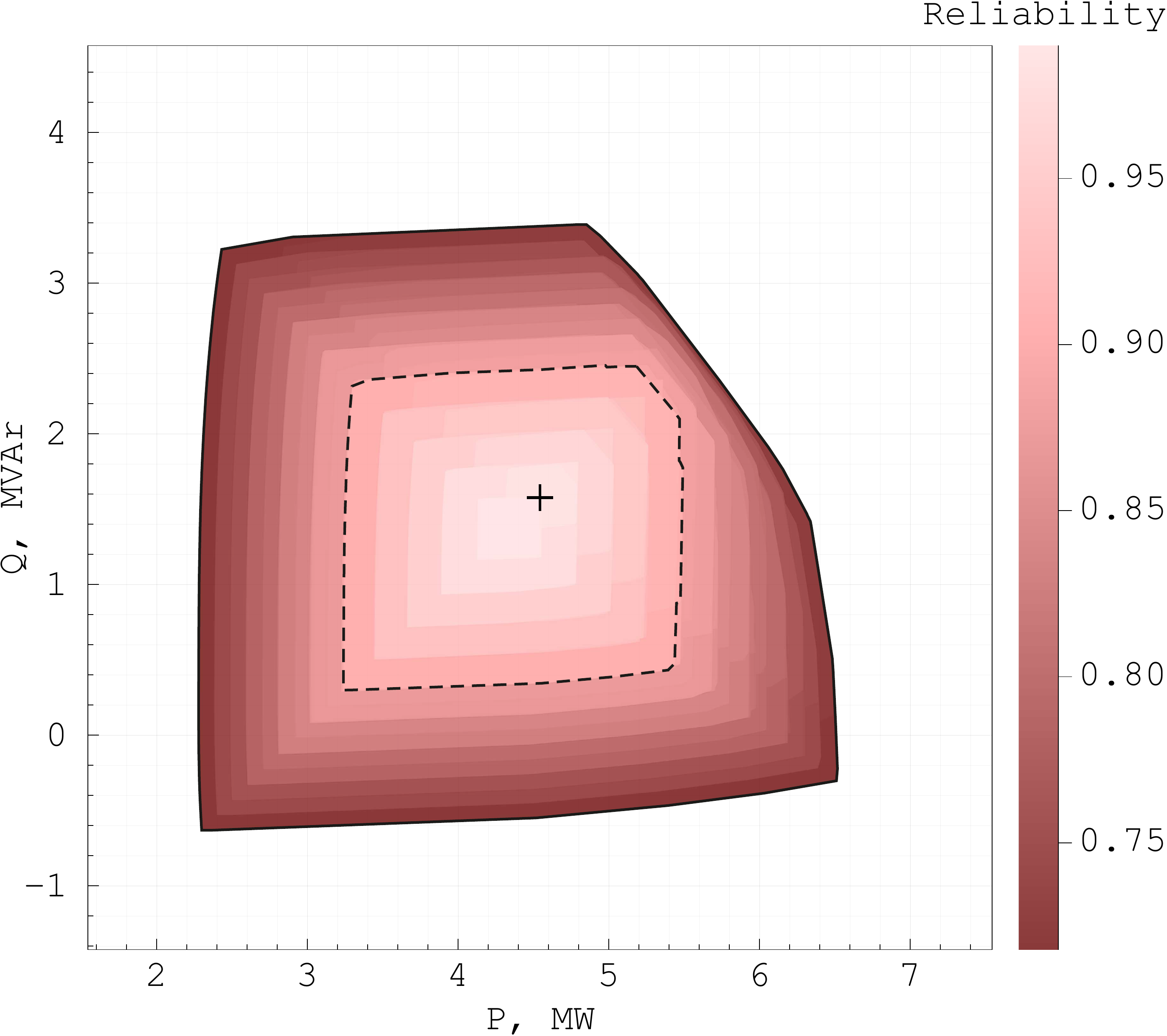}}
    \subfigure(b){\includegraphics[width=0.47\linewidth]{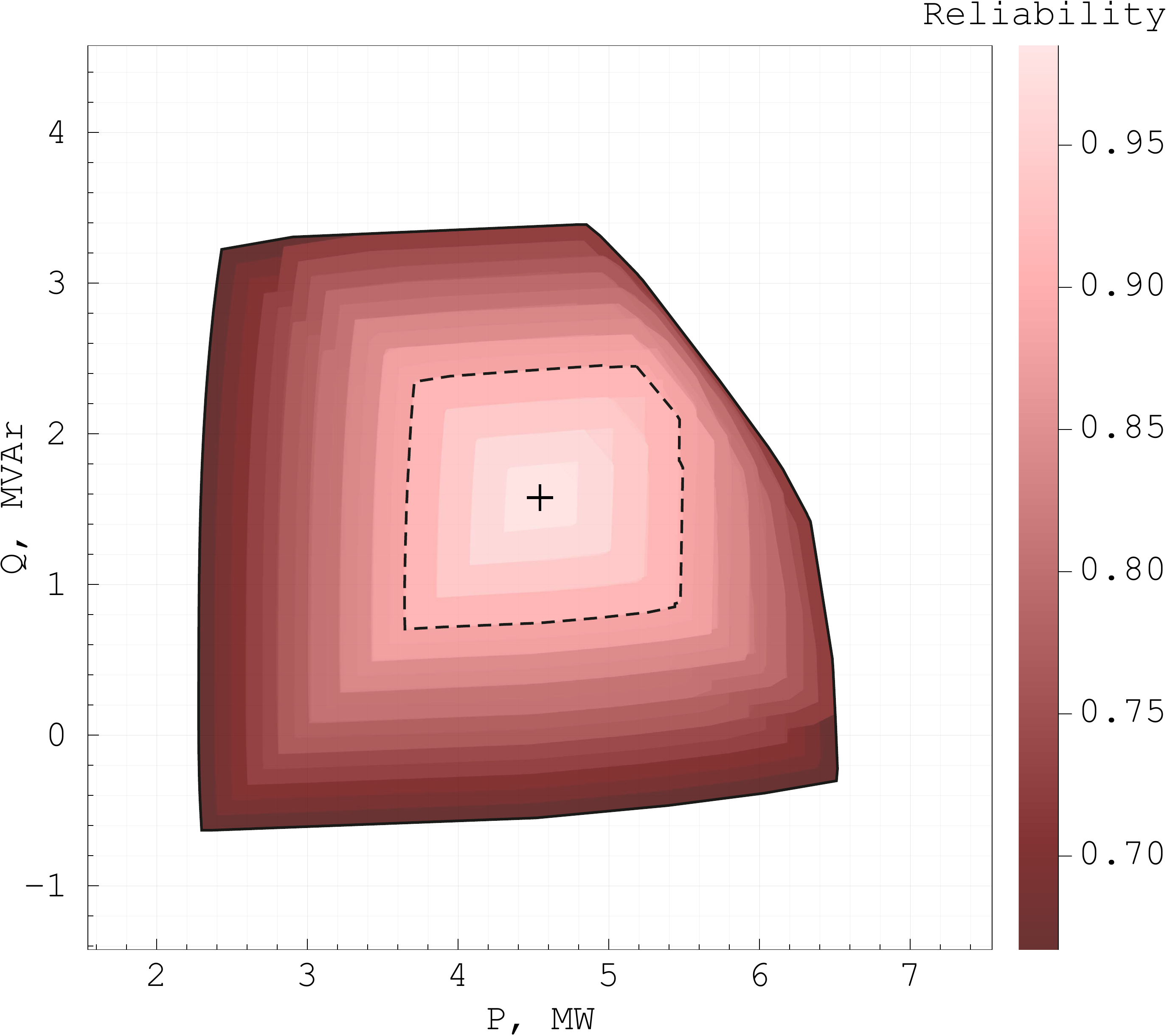}}
    \caption{Reliability-based segmentation of the 124-bus network P-Q flexibility area: (a) FU at bus 20 is considered the most reliable one with $R_{20}=0.99$; (b) reliability estimation for the unit decreased down to $R_{20}=0.92$. The displayed flexibility areas correspond to the network configuration with the power supply via feeder 7-1. The red color scheme indicates the reliability associated with each of the segments. The dashed outlines contain the unions of the flexibility segments with reliability greater than $0.9$.}
    \label{Fig: Griffin - arranged segments}
\end{figure*}

The proposed reliability-based segmentation method is used to produce all possible combinations of FUs' P-Q capabilities and arrange them by their reliability in decreasing order, $\mathcal{R}_1\geq\mathcal{R}_2 \geq ... \geq \min_{\text{seg}\in S}(\mathcal{R}_{\text{seg}})$. Note that estimating flexibility for all possible combinations of FUs is a complex combinatorial problem that can be impractical to solve. In the case of 10 FUs, there are only 1023 possible combinations to simulate. In larger case studies, the problem can be simplified by limiting the number of segments to consider, e.g., DSO might only be interested in a handful of reliability levels to analyse (e.g., greater than 80\%, 95\%, etc.). The reliability-based segmentation for the networks' flexibility area is visualised in Fig.~\ref{Fig: Griffin - arranged segments}(a). The central segments indicate reliable operating points that can be reached by activating a few FUs with the highest reliabilities. The reliability of activating more units decreases as operating points become closer to the boundary of the aggregated flexibility area. DSO can use this segmentation to analyse and manage the flexibility available in the network. For example, DSO might plan to provide flexibility services whose estimated reliability exceeds $0.9$. Thus, only a fraction of the segments can be considered for network operation. These segments are depicted in Fig.~\ref{Fig: Griffin - arranged segments} by a dashed outline. The area with reliability exceeding $0.9$ contains much fewer operating points than the entire flexibility area. It follows that the aggregated flexibility can be significantly overestimated without considering reliability constraints.

The estimated reliabilities of FUs can change throughout the network operation. Such changes can have a major impact on the aggregated flexibility area and its segmentation. To illustrate this, it is assumed that the reliability associated with unit 20, $R_{20}$, drops from $0.99$ to $0.92$. I.e., the most reliable unit is turned into the least reliable one. Then, the reliability analysis of FUs combinations is repeated. The modified reliability-based segmentation is displayed in Fig.~\ref{Fig: Griffin - arranged segments}(b). Even though the aggregated flexibility area stays the same, its components change drastically. Specifically, the reliabilities of the flexibility area segments decrease. The union of the segments with reliability exceeding $0.9$ (the dashed outline) shrinks compared to the simulation presented in Fig.~\ref{Fig: Griffin - arranged segments}(a). Thus, the proposed flexibility area segmentation framework captures the effects of changes in reliability estimation and network parameters. DSO can update the network's flexibility segmentation regularly to manage its flexibility services provided.

\section{Conclusion}
This paper highlights the need for a thorough analysis of the flexibility inherent in ADNs and the reliability of flexibility services. Specifically, it is required to characterise flexibility by explicitly considering the contributions and reliability of each FU. As illustrated by the simulations for a real distribution network, the flexibility of a network can change significantly depending on the network configuration, technical limits, and reliability of FUs. By capturing these effects, the proposed flexibility area segmentation framework enables DSOs to manage the provision of ADN flexibility services, e.g., by increasing the reliability of the services whereas reducing the number of dispatched units and control actions. Moreover, the obtained segmentation can also be used as an input for risk assessment of distribution networks and optimisation problems at the transmission level, e.g., to consider solutions for TSO-DSO coordination with predefined levels of reliability.

\bibliographystyle{IEEEtran}
\bibliography{references.bib}

\end{document}